# Inhibition of the phase conjugation of orbital angular momentum superpositions in cylindrical vector beams during stimulated Brillouin scattering

Jean-François Bisson

Département de physique et d'astronomie, Université de Moncton

Jean-francois.bisson@umoncton.ca

Phase conjugation with stimulated Brillouin scattering can be used to correct wavefront aberrations in high-power laser systems. Here, we consider such process with cylindrical vector lasers beams. By using a vectorial approach and a modal decomposition, we obtain a differential matrix equation that enables the calculation of the structure of Stokes vector eigenmodes and their Brillouin gain. The emphasis is put on the mode with the largest gain, which contributes the most to the Stokes beam. We show that the phase conjugation of orbital angular momentum in cylindrical vector beams is prevented from happening almost everywhere on the higher-order Poincaré sphere. This phenomenon is traced to the tendency of the Stokes beam to acquire a polarization structure similar to the incident pump beam, which constraints the possible combinations of topological charges of the two orbital angular momentum modes making up the Stokes cylindrical vector beams. However, near the poles of the higher-order Poincaré sphere, the Stokes mode with the largest gain acquires an additional helical phase factor that produces the conjugated topological charge for the stronger circularly polarized component without affecting the polarization structure. Experimental results are found to agree with our theoretical predictions.

# I. INTRODUCTION

Phase conjugation (PC) in optics is a process whereby an optical wave, after interacting with a non-linear medium, acquires a structure that is the complex conjugate of the incident wave, namely the amplitude and the polarization state of the incident wave are replaced by their respective complex conjugate after reflection [1]. PC can be achieved with various methods such as four-wave mixing [2], stimulated parametric down conversion [3] and stimulated Brillouin scattering (SBS) [4,5]. The latter, discovered by Zeldovich [6], has found application in high-power laser systems for the correction of aberrated wavefronts originating from thermal effects taking place inside the active material [7]. In this case, the SBS cell is used as a laser mirror which redirects a phase-conjugate beam into the active material, wherein the wavefront distortions are compensated after the second pass. Now, this PC process is an imperfect one because the polarization state of the reflected wave is usually the same as that of the incoming wave for homogeneously polarized laser beams [4,5]. Hence, reflection with a SBS mirror is generally considered a *scalar* PC process, in contrast to *vectorial* PC [8,9,10], which conjugates the polarization state as well.

Now the question arises as what happens when SBS takes place with a beam of light having both spatially inhomogeneous polarization and phase distributions? Such beams are called vector beams, a rapidly expanding research topic that has fueled numerous applications [11,12,13]. Of particular interest are rotation-invariant solutions, called cylindrical vector beams (CVBs), such as:

$$\vec{u} = \sin(\theta/2)\exp(i\alpha)LG_0^{-1}(\vec{r})\hat{e}_L + \cos(\theta/2)LG_0^{1}(\vec{r})\hat{e}_R \quad , \tag{1}$$

where $LG_{p=0}^{m=\pm 1}$ are Laguerre-Gauss modes of radial and azimuthal indices $p=0$ and $m=\pm 1$ with helical wavefronts of the form $\exp(\pm i\varphi)$ ; $\hat{e}_R = \frac{1}{\sqrt{2}}(\vec{e}_x - i\vec{e}_y)$ and $\hat{e}_L = \frac{1}{\sqrt{2}}(\vec{e}_x + i\vec{e}_y)$ are circular polarization states of opposite handedness.[1] CVBs with high energy have received attention, notably to achieve large values of longitudinal electro-magnetic fields for charge acceleration [14,15]. Such beam can also be efficiently amplified by SBS [16].

CVB states can be mapped onto a sphere of unit radius, with polar and azimuthal coordinates $\theta$ and $\alpha$, called the higher-order Poincaré sphere (HOPS) [17]. Examples of the structure of a CVB state of eq. (1) at various locations of the HOPS are shown in Fig. 1. At the north pole of the HOPS, $\theta=0º$, the polarization is right circular; the eccentricity of the elliptic polarization increases as $\theta$ increases along a line of longitude of the HOPS to reach rectilinear polarization at the equator ($\theta=90º$), becomes left-handed at higher $\theta$ values to reach left-circular polarization at the south pole ($\theta=180º$).

Now, the usual prescription used in scalar PC, whereby the complex conjugate of the phase is taken while the polarization remains unchanged, is unapplicable for CVBs because the phase of such beams is affected by both the helical structure of the wavefront and by the changing polarization state along $\varphi$. For instance, if the weight of each circularly-polarized component of the pump beam remains the same after reflection from a SBS mirror, while the sign of the topological charge is reversed, then the CVB loses its rotational symmetry and the polarization ellipse rotates in the opposite direction with $\varphi$, as shown in Fig. 2. Indeed, one can show that the phase of a CVB has contributions from both the helical wavefronts and from the geometrical phase arising from the changing polarization state with the azimuthal position $\varphi$ (see Supplemental Material [18]). Therefore, the phase and the polarization states in a CVB are entangled quantities and the pure scalar approach to describe SBS with such a beam is inadequate.

Vectorial SBS theories do exist. They often have different meanings or purposes. For instance, there is vectorial PC with SBS [8,9], wherein a beam having both wavefront and polarization distortions is split into two separate orthogonally polarized beams that are then focused into a SBS cell where the two beams overlap. The nonlinear interaction inside the overlapping region allows the phase of the two beams to be locked [19]. The combination of the SBS reflected beams then allows one to achieve vector PC, where both the wavefront and polarization states are conjugated. Such experiment can be understood with the scalar theory. Second, one may analyze the polarization properties of a SBS beam as a function of the polarization state of the pump beam and the optical properties of the SBS medium. The behavior of SBS of inhomogeneously polarized beams in free space was studied at the early stages of the discovery of this process: the polarization inhomogeneity was produced randomly by using birefringent materials of poor optical quality [20] or by etching such birefringent materials to produce differences in the optical path in the transverse direction [21]. But the results obtained cannot readily be transposed to vector beams, which have a well-defined, deterministic structure in the polarization distribution. The polarization properties of SBS were also studied for birefringent single mode fibers [22,23]. However, since single mode fibers support only one transverse mode, this modeling needs to be adapted to correctly address the description of SBS of CVBs and how the modes with different topological charges are modified (or not) by the nonlinear SBS interaction.

---

[1] We use the same system of coordinates for both the incident and reflected waves, e.g., an incident $\exp(im\varphi)\hat{e}_R$ state remains $\exp(im\varphi)\hat{e}_R$ after reflection from a classical mirror even though the reversal of the $k$ vector changes a right-circular into a left-circular state. We also use the convention exp(-i$\omega$t).

Here, the problem of the phase conjugation of CVBs, by virtue of being a superposition of orthogonally polarized OAM modes, calls for a modal analysis. Moore and Boyd [24] did use a spatial mode decomposition approach, but they did so only in the scalar framework with homogeneously polarized Hermite-Gauss modes. We report, for the first time to our knowledge, a vectorial analysis that includes a modal decomposition for the description SBS of CVBs. Our analysis considers both the inhomogeneous polarization and the superposition of states of different topological charges by implementing a modal decomposition within a vectorial description of SBS. We also show experimental results that support our theoretical findings.

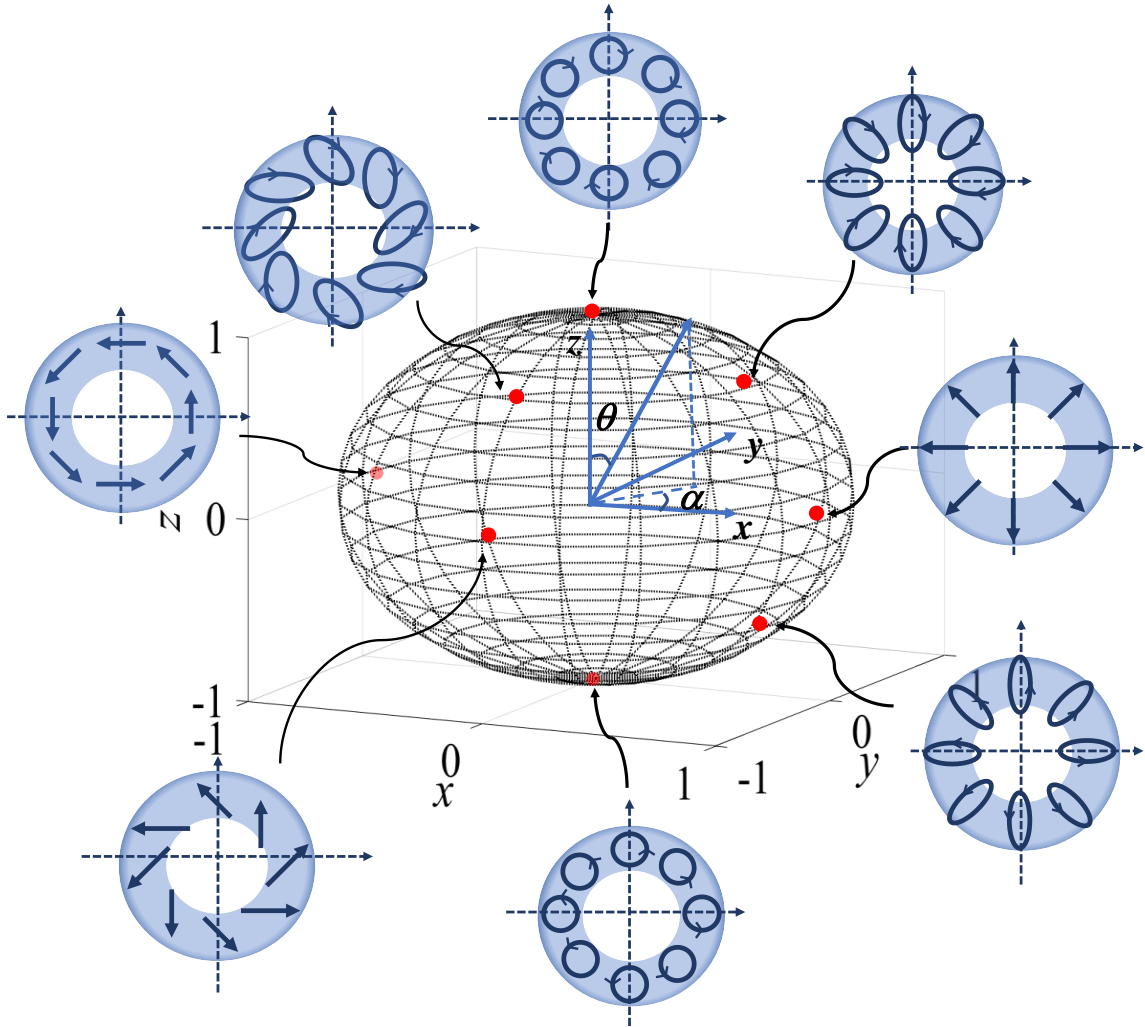

Fig. 1 Diagram of the CVB described by eq. (1) on the HOPS. The phase distribution is indicated by the position of the arrow on each ellipse of polarization at a given time as a function of $\varphi$.

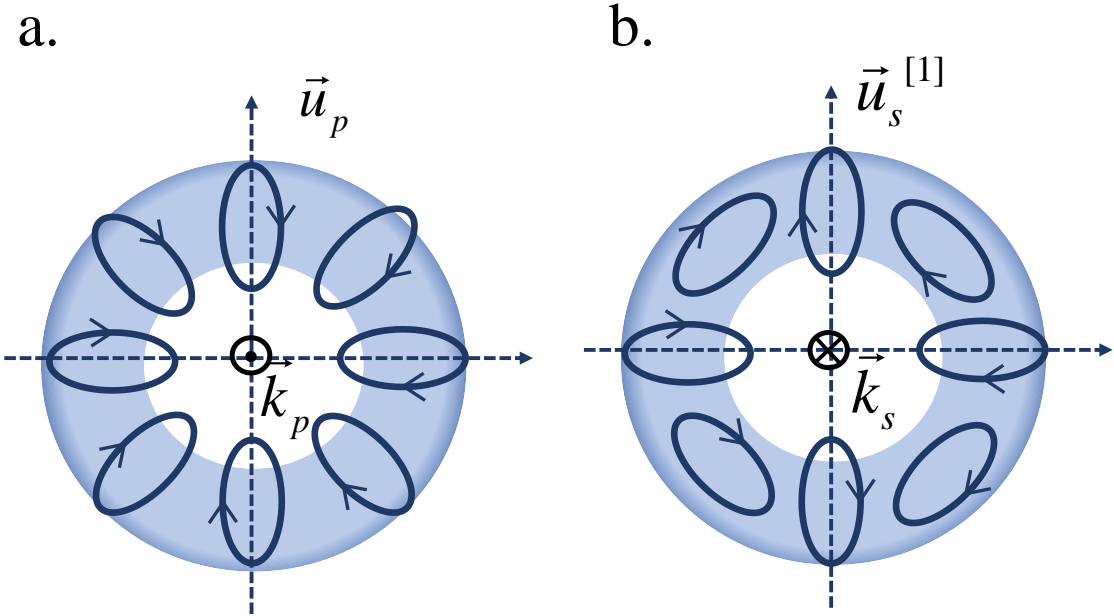

Fig. 2. Polarization distribution of a) pump and b) hypothetical Stokes beams, assuming that the topological charges change sign while each circular polarization state remains unchanged at the reflection, $\vec{u}_s^{[1]} = \sin(\theta/2)\exp(i\varphi)\hat{e}_L + \cos(\theta/2)\exp(-i\varphi)\hat{e}_R$. This scenario, where the Stokes CVB changes its polarization structure in SBS reflection, breaks the rotational symmetry and is not favored according to our modeling and experimental results.

This paper is structured as follows. In section II, we present a theory of the SBS process that takes into account the vectorial structure and the modal content of the pump and Stokes beams. One key point is the fact that the pump and Stokes beams need not have the same local polarization state, in contrast to the scalar theory. This affects the gain of the Stokes wave via the reduced contrast of the density modulation arising from the interference pattern of the pump and Stokes waves of different polarization states. We use a modal decomposition of the two beams, with our attention turned to the azimuthal dependence of the electric field vector of the CVB. In the undepleted pump approximation, we show that the SBS interaction only couples OAM beams of topological charges $n$ and $n$-2 for the left and right circular polarization states respectively, via a first-order matrix differential equation. The eigenvectors of this matrix equation determine the mode structure. The mode with the largest eigenvalue has the highest gain and thus bears particular significance in the prediction of the experimentally observed beam structure. We show that the conjugation of the OAM, i.e., the sign reversal of topological charge $m$ in eq.

(1), of CVBs is prevented from taking place in most regions of the HOPS. However, near the poles of the HOPS, where the CVB approaches the structure of a scalar beam, the Stokes beam is found to acquire a global helical phase factor of exp(-2i$\varphi$), that does produce the conjugated topological charge for the stronger component and a topological charge of *m*=-3 for the smaller component of eq. (1). The effect of a slight departure from a perfectly axisymmetric beam is also analyzed. It is shown to lift the gain degeneracy generally observed for a pump beam with a pure vortex mode for topological charges *m* = -1, 0, and 1 of the Stokes beam, in favor of the conjugated OAM.

In section III, experimental results are shown for a CVB, represented by eq. (1), interacting with a Brillouin medium. Using waveplates and a vortex plate, CVBs are created at various locations of the HOPS. The topological charge of each circularly polarized component of the Stokes beam is separately analyzed using a circular polarizer and a shearing interferometer. The Stokes beam is found to have the same topological charge as the pump beam for each circular polarization away from the pole ($\theta$=0º); however, as $\theta$ approaches 0º, the topological charges of left and right circular polarizations gradually shift from ($m_{left}$= -1, $m_{right}$= 1) to ($m_{left}$= 2, $m_{right}$= 0) to ($m_{left}$= -3, $m_{right}$= -1), at which point most of the power is in the right-circular mode and phase conjugation of the OAM takes place. Our results agree with our theoretical model. We wrap up this paper in section IV with a discussion and a conclusion.

## II. THEORY

The theory of SBS with scalar beams is well known [4,5]. We outline it and adapt it to the vectorial nature of our beam. Then we make a modal decomposition following a procedure outlined by Moore and Boyd [24]. One starts with the wave equation for the electric field:

$$\nabla^2 \vec{E} - \frac{\kappa}{c^2} \frac{\partial^2 \vec{E}}{\partial t^2} = \frac{1}{\varepsilon_0 c^2} \frac{\partial^2 \vec{P}^{NL}}{\partial t^2} ,\qquad(2)$$

where $\kappa$ is the dielectric constant and $\vec{P}^{NL}$ is a third-order nonlinear polarization term driving the SBS process. We assume the field is described by two counter-propagating, monochromatic pump and Stokes waves with slowly varying envelopes:

$$\vec{E} = \vec{E}_p(\vec{r})\exp(-i(kz+\omega_p t)) + \vec{E}_s(\vec{r})\exp(i(kz-\omega_s t)). \tag{3}$$

Their interference pattern inside the Brillouin medium produces, through electrostriction, a density modulation receding from the pump. The pump wave can be converted into a Stokes wave by Bragg reflection from the resulting permittivity modulation, thereby reinforcing the interference pattern, producing the SBS phenomenon. Inserting eq. (3) into eq. (2), and applying the paraxial approximation, i.e.,

$$\vec{E}_p = \begin{pmatrix} E_{p,x} & E_{p,y} & 0 \end{pmatrix}^T, \tag{4a}$$

$$\vec{E}_s = \begin{pmatrix} E_{s,x} & E_{s,y} & 0 \end{pmatrix}^T, \tag{4b}$$

$$\frac{\partial^2 \vec{E}_{p(s)}}{\partial z^2} \ll k\frac{\partial \vec{E}_{p(s)}}{\partial z}, \frac{\partial^2 \vec{E}_{p(s)}}{\partial z^2} \ll \frac{\partial^2 \vec{E}_{p(s)}}{\partial x^2}, \frac{\partial^2 \vec{E}_{p(s)}}{\partial z^2} \ll \frac{\partial^2 \vec{E}_{p(s)}}{\partial y^2}, \tag{4c}$$

and assuming steady state, we obtain:

$$\nabla_\perp^2 \vec{E}_p - 2ik\frac{\partial \vec{E}_p}{\partial z} = -\frac{\omega^2}{\varepsilon_0 c^2}\vec{P}_{-ikz}^{NL} \tag{5a}$$

$$\nabla_\perp^2 \vec{E}_s + 2ik\frac{\partial \vec{E}_s}{\partial z} = -\frac{\omega^2}{\varepsilon_0 c^2}\vec{P}_{ikz}^{NL}, \tag{5b}$$

where $\omega_p \approx \omega_s = \omega$, $\nabla_\perp^2 \equiv \frac{\partial^2}{\partial x^2}+\frac{\partial^2}{\partial y^2}$ and $\vec{P}_{\pm ikz}^{NL}$ are the portions of the nonlinear polarization phase matched to the pump (-i$kz$) and Stokes (+i$kz$) waves. For scalar beams, interference terms $E_p^* E_s$ and $E_p E_s^*$ are responsible for the intensity modulation; they produce polarization contributions phase-matched to the pump and Stokes waves respectively given by [4]:

$$P_{-ikz}^{NL} = \varepsilon_0 \chi_{SBS}^{(3)}\left(E_s^* \cdot E_p\right)E_s, \tag{6a}$$

$$P_{ikz}^{NL} = -\varepsilon_0 \chi_{SBS}^{(3)}\left(E_p^* \cdot E_s\right)E_p. \tag{6b}$$

Now, for vector beams, the pump and Stokes wave need not have the same polarization state. In the extreme case where $\vec{E}_p$ and $\vec{E}_s$ are orthogonally polarized, the density modulation arising from electrostriction, responsible for SBS, disappears. The modulation terms must be modified to $\vec{E}_p^* \cdot \vec{E}_s$ and $\vec{E}_s^* \cdot \vec{E}_p$, as exposed in Ref. [22-23]. Eqs. (6) become instead:

$$\vec{P}_{-ikz}^{NL} = \varepsilon_0 \chi_{SBS}^{(3)} \left( \vec{E}_s^* \cdot \vec{E}_p \right) \vec{E}_s, \tag{7a}$$

$$\vec{P}_{+ikz}^{NL} = -\varepsilon_0 \chi_{SBS}^{(3)} \left( \vec{E}_p^* \cdot \vec{E}_s \right) \vec{E}_p. \tag{7b}$$

Next, we follow a procedure outlined by Moore and Boyd [24], where we use a modal decomposition of the pump and Stokes fields:

$$\vec{E}_p = \left( \sum_j a_j(z) A_j(r,\varphi,z) \quad \sum_j a'_j(z) A_j(r,\varphi,z) \right)^T, \tag{8a}$$

$$\vec{E}_s = \left( \sum_j b_j(z) B_j(r,\varphi,z) \quad \sum_j b'_j(z) B_j(r,\varphi,z) \right)^T, \tag{8b}$$

where $\{A_j\}$ and $\{B_j\}$ form an orthonormal set of eigenmodes for the pump and Stokes waves, and $a_j$, $a'_j$, $b_j$, and $b'_j$ are complex coefficients. Substituting eqs. (8) into eqs. (7) and then into eqs. (5) and using the fact that the functions $\{A_j\}$ and $\{B_j\}$ are solutions of the homogeneous wave equation, we find:

$$\begin{pmatrix} \sum_j A_j \dfrac{\partial a_\alpha}{\partial z} \\ \sum_j A_j \dfrac{\partial a'_j}{\partial z} \end{pmatrix} = g \sum_{j,l,m} \begin{pmatrix} \left( a_l b^*_j b_m + a'_l b'^*_j b_m \right) A_l B_j^* B_m \\ \left( a_l b^*_j b'_m + a'_l b'^*_j b'_m \right) A_l B_j^* B_m \end{pmatrix}, \tag{9a}$$

$$\begin{pmatrix} \sum_j B_j \dfrac{\partial b_\alpha}{\partial z} \\ \sum_j B_j \dfrac{\partial b'_\alpha}{\partial z} \end{pmatrix} = g \sum_{j,l,m} \begin{pmatrix} \left( b_l a^*_j a_m + b'_l a'^*_j a_m \right) B_l A_j^* A_m \\ \left( b_l a^*_j a'_m + b'_l a'^*_j a'_m \right) B_l A_j^* A_m \end{pmatrix}, \tag{9b}$$

where $g \equiv -i \dfrac{\omega^2 \chi_{SBS}^{(3)}}{2kc^2}$ is a gain factor. Then we multiply each eq. (9) by one basis function $A_n^*$ or $B_n^*$, integrate over transverse coordinates and, using orthonormality, we find:

$$\begin{pmatrix} \dfrac{\partial a_n}{\partial z} \\ \dfrac{\partial a'_n}{\partial z} \end{pmatrix} = g \sum_{j,l,m} \begin{pmatrix} a_l b^*_j b_m + a'_l b'^*_j b_m \\ a_l b^*_j b'_m + a'_l b'^*_j b'_m \end{pmatrix} \xi_{p,jlmn} , \qquad (10a)$$

and

$$\begin{pmatrix} \dfrac{\partial b_n}{\partial z} \\ \dfrac{\partial b'_n}{\partial z} \end{pmatrix} = g \sum_{j,l,m} \begin{pmatrix} b_l a^*_j a_m + b'_l a'^*_j a_m \\ b_l a^*_j a'_m + b'_l a'^*_j a'_m \end{pmatrix} \xi_{s,jlmn} , \qquad (10b)$$

where:

$$\xi_{p,jlmn}(z) = \int A_l B_j^* B_m A_n^* \, d^2\vec{r} \qquad (11a)$$

and

$$\xi_{s,jlmn}(z) = \int B_l A_j^* A_m B_n^* \, d^2\vec{r} \qquad (11b)$$

are overlap integrals of the modes over transverse coordinates. For the calculation of $\xi_{p,jlmn}$ and $\xi_{s,jlmn}$, it is convenient to choose eigenmodes with well-defined topological charges $m$ such as:

$$A_{mp}(r,\varphi,z) = C_{mp}(r,z)\exp(im\varphi),$$
(12a)

and choose the $\{B_i\}$ identical to the $\{A_i\}$ modes:

$$B_{mp}(r,\varphi,z) = A_{mp}(r,\varphi,z). \qquad (12b)$$

Then, $\xi_{p,jlmn} = \xi_{s,jlmn} \equiv \xi_{jlmn}$. Next, we focus our attention on the transformation of OAM by the SBS process: indices $j$, $l$, $m$, and $n$ may then be interpreted as topological charges. In such case, one can see that:

$$\xi_{jlmn} = 0 \tag{13}$$

unless:

$$j - l - m + n = 0. \tag{14}$$

Next, the pump beam in eq. (1) is given, in the $\{\hat{e}_L, \hat{e}_R\}$ circular polarization basis, by:

$$\vec{E}_p \sim \begin{pmatrix} a_{-1} A_{-1} \\ a'_1 A_1 \end{pmatrix}_{\{\hat{e}_L, \hat{e}_R\}}, \tag{15}$$

where

$$a_{-1} = \sin(\theta/2)\exp(-i\alpha); \quad a_n = 0, \quad \text{for } n \neq -1$$
$$a'_1 = \cos(\theta/2); \quad a'_n = 0, \quad \text{for } n \neq 1 \tag{16}$$

Finally, we assume that the undepleted pump approximation is valid, which implies that the coefficients $a_n$ and $a'_n$ are constant. Inserting eqs. (15-16) into eq.(10b) and using eqs. (13-14), we find:

$$\frac{\partial}{\partial z}\begin{pmatrix} b_{n-2} \\ b'_n \end{pmatrix} = M \begin{pmatrix} b_{n-2} \\ b'_n \end{pmatrix}, \tag{17}$$

where $M$ is a two-by-two matrix given by:

$$M = \begin{pmatrix} \sin^2\left(\frac{\theta}{2}\right)\xi_{11} & \sin\left(\frac{\theta}{2}\right)\cos\left(\frac{\theta}{2}\right)\exp(-i\alpha)\xi_{12} \\ \sin\left(\frac{\theta}{2}\right)\cos\left(\frac{\theta}{2}\right)\exp(i\alpha)\xi_{21} & \cos^2\left(\frac{\theta}{2}\right)\xi_{22} \end{pmatrix}, \tag{18}$$

and

$$\xi_{11} \equiv \xi_{-1,n-2,-1,n-2}, \tag{19a}$$

$$\xi_{12} \equiv \xi_{1,n,-1,n-2}, \tag{19b}$$

$$\xi_{21} \equiv \xi_{-1,n-2,1,n}, \tag{19c}$$

$$\xi_{22} \equiv \xi_{1,n,1,n}. \tag{19d}$$

The diagonalization of eq. (18) allows one to find the Stokes eigenmodes for each value of $n$; among them, the one with the highest gain, i.e., the one with the largest eigenvalue, is the one that should emerge and most likely to be observed experimentally. The main conclusions of this paper are based on eqs. (17-18). The inspection of matrix $M$ reveals that:

1. The vector modes of the Stokes beam contain one circular polarization component with a topological charge $m=n$ and an orthogonal circular polarization component with a topological charge $m=n-2$.

2. Each vector mode $\vec{E}_s = \left(b_{n-2}\exp(i(n-2)\varphi) \quad b_n\exp(in\varphi)\right)^T_{\{\hat{e}_L,\hat{e}_R\}}$ behaves independently in this model, i.e., there is no crosstalk with other vector beam solutions with different $n$ values.

3. Matrix $M$ is Hermitian since $\xi_{ij}$ values are real and $\xi_{12}=\xi_{21}$. Thus, the eigenvectors are orthogonal and the eigenvalues are real. The eigenvector with the larger eigenvalue has higher gain and is more likely to be experimentally observed.

4. The presence of $\xi_{ij}$ in $M$ makes its eigenvectors in general different from $\vec{E}_p \sim \left(a_{-1}A_{-1} \quad a'_1 A_1\right)^T_{\{\hat{e}_L,\hat{e}_R\}}$ with the coefficients $a_1$ and $a'_1$ given by eq. (16). Hence, the polarization states of the Stokes beam need not be identical to the pump beam, in contrast to scalar beams (see ref. [18], Table S1).

The largest eigenvalues of eq. (18) for different Stokes modes $\vec{E}_{s n, n-2}$ and different $\theta$ values, for $\alpha=0$, of the pump are shown in Table 1 for $A_{\pm 1} = LG_{p=0}^{m=\pm 1}$ beams. One can see that the Stokes mode $\vec{E}_{s,-1,1} = \left(b_{-1}\exp(-i\varphi) \quad b_1\exp(i\varphi)\right)^T$ has the largest gain near the equator of the HOPS but the Stokes vector mode $\vec{E}_{s,-2,0}$ and $\vec{E}_{s,-3,-1}$ become increasingly competitive as the pump CVB approaches the pole of the HOPS. At that point, the beam becomes almost scalar and the fundamental Gaussian mode with azimuthal index $m=0$ and the conjugate mode, $m=-1$, become as strong as $m=1$, in conformity to previous reports of phase conjugation of scalar vortex beams [25-26].

Table 1. Calculated normalized maximum eigenvalue of eigen-equations (17-18) for Stokes CVB with topological charges (*n-2, n*) in $\hat{e}_L$ and $\hat{e}_R$ polarization states. Values in bold designate the mode with highest gain.

| $\theta$ (deg.) | (*n-2, n*) = (-4, -2) | (-3, -1) | (-2, 0) | (-1, 1) | (0, 2) |
|---|---|---|---|---|---|
| 90 | 0.51 | 0.71 | 0.80 | **1.00** | 0.80 |
| 82 | 0.54 | 0.74 | 0.82 | **1.00** | 0.78 |
| 74 | 0.57 | 0.78 | 0.84 | **1.00** | 0.77 |
| 66 | 0.60 | 0.82 | 0.87 | **1.00** | 0.76 |
| 58 | 0.63 | 0.86 | 0.89 | **1.00** | 0.75 |
| 50 | 0.66 | 0.89 | 0.92 | **1.00** | 0.75 |
| 42 | 0.69 | 0.92 | 0.94 | **1.00** | 0.74 |
| 34 | 0.71 | 0.95 | 0.96 | **1.00** | 0.75 |
| 26 | 0.72 | 0.97 | 0.98 | **1.00** | 0.75 |
| 18 | 0.74 | 0.98 | 0.99 | **1.00** | 0.75 |
| 10 | 0.75 | **1.00** | **1.00** | **1.00** | 0.75 |
| 2 | 0.75 | **1.00** | **1.00** | **1.00** | 0.75 |

However, we shall see section III that we experimentally observed a domination of the mode (-3,-1), near the poles of the HOPS, in contrast with the predictions shown in Table 1 (Cf. section III). This mode corresponds to the conjugation of the OAM for the stronger circular component near $\theta=0°$. We hypothesize that this originates from a pump wave that was not perfectly axisymmetric. This hypothesis was tested in our model by building a set of orthogonal eigenmodes as:

$$A_{m,\varepsilon} = LG_m + \varepsilon \operatorname{sign}(m) LG_{-m} \quad (20a)$$

and

$$B_{m,\varepsilon} = LG_m - \varepsilon \operatorname{sign}(m) LG_{-m}, \quad (20b)$$

where parameter $\varepsilon$ is a small real number. The parameter $\varepsilon$ introduces an asymmetry in the intensity pattern. Crucially, $A_{m,\varepsilon}$ and $A_{-m,\varepsilon}$ no longer have the same intensity distribution. Moreover, $B_{-m,\varepsilon} = A_{m,\varepsilon}^*$; thus, $A_{m,\varepsilon}^*$ and $B_{-m,\varepsilon}$ have the same intensity profile. This eliminates the equality of Brillouin gain theoretically predicted for LG modes of opposite $m$ values. The $\xi$ values for $\varepsilon>0$ can easily be calculated from those for $\varepsilon=0$ (Cf. Ref. [18], section 1.2). The calculation of the eigenvalues of the dominant eigenvector is shown in Table 2 for the case $\varepsilon=0.15$, which qualitatively matches our experimental results. Whereas the situation remains unchanged near the equator (no conjugation of OAM), a shift of the topological of the dominant mode from $n=1$ to $n=-1$ is found, while the weaker state of opposite circular polarization acquires a topological charge of $n=-3$. This result is consistent with previous reports showing that an aberrated pump beam produced a reflected Stokes with a better fidelity [6, 26-27].

Table 2. Calculated normalized maximum eigenvalues for non-axisymmetric eigenfunctions of eq. (20) with $\varepsilon=0.15$ for Stokes CVB with topological charges ($n-2$, $n$) in $\hat{e}_L$ and $\hat{e}_R$ polarization states. Values in bold character designate the mode with highest gain.

| $\theta$ (deg.) | $(n-2, n) =$ $(-4, -2)$ | $(-3, -1)$ | $(-2, 0)$ | $(-1, 1)$ | $(0, 2)$ |
|---|---|---|---|---|---|
| 90 | 0.48 | 0.68 | 0.75 | **0.92** | 0.75 |
| 82 | 0.51 | 0.72 | 0.77 | **0.92** | 0.73 |
| 74 | 0.54 | 0.76 | 0.79 | **0.92** | 0.72 |
| 66 | 0.57 | 0.80 | 0.82 | **0.92** | 0.71 |
| 58 | 0.60 | 0.84 | 0.84 | **0.92** | 0.71 |
| 50 | 0.63 | 0.88 | 0.87 | **0.92** | 0.71 |
| 42 | 0.66 | 0.91 | 0.89 | **0.92** | 0.71 |
| 34 | 0.68 | **0.94** | 0.92 | 0.92 | 0.71 |
| 26 | 0.69 | **0.97** | 0.93 | 0.92 | 0.71 |
| 18 | 0.71 | **0.98** | 0.95 | 0.92 | 0.72 |
| 10 | 0.72 | **1.00** | 0.95 | 0.92 | 0.72 |
| 2 | 0.72 | **1.00** | 0.96 | 0.92 | 0.72 |

## III. EXPERIMENTS

A $Nd^{3+}$-doped $Y_3Al_5O_{12}$ single-mode laser delivering linearly-polarized, 3-ns pulses at wavelength $\lambda$=1064 nm is prepared into an elliptical polarization state using an adjustable half-wave plate (HWP) followed by a fixed quarter-wave plate (QWP) with its fast axis in the horizontal direction. The laser pulses then pass through an inhomogeneous HWP, also known as a $q$-plate [28], wherein the orientation $\theta_{\lambda/2}$ of the fast axis with respect to the horizontal axis rotates with the azimuthal angle as $\theta_{\lambda/2}=\varphi/2$ to create a CVB described by eq. (1) with fixed $\alpha$=0º and variable $\theta$ value depending on the orientation of the half-wave plate. The laser pulses are then focused into a SBS cell containing liquid $CCl_4$ with a converging lens with focal length $f$=10 cm. The incident pulse energy is $E$=3.5 mJ; the SBS threshold was previously measured to be around $E_{th,SBS}$=1 mJ. The reflected Stokes beam is redirected towards a diagnostic apparatus using a wedge (W) fused silica window and a dielectric Bragg mirror (M); the incidence angle on these elements is less than 10º in order to minimize changes of the polarization state, Fig. 3.

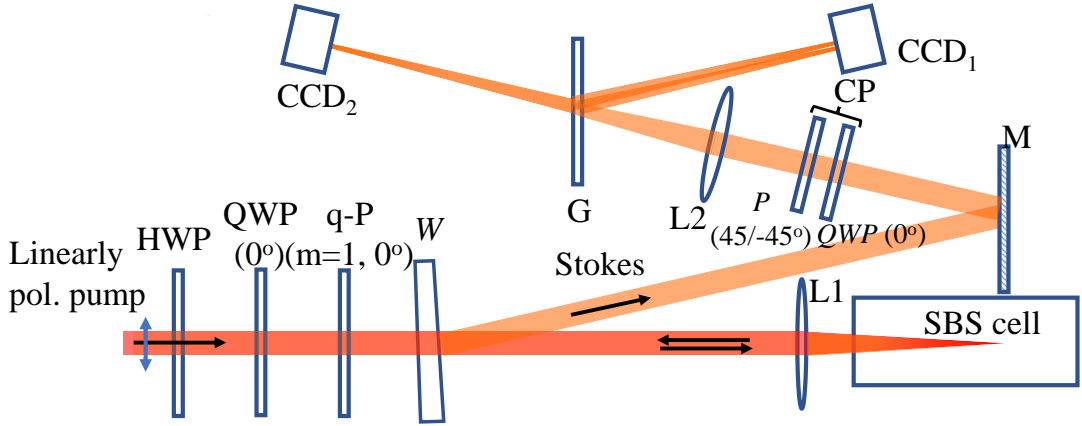

Fig. 3. Experimental set-up showing the preparation of the CVB with the HWP-QWP-q-plate focused into a SBS cell with lens L1. The diagnostic of the Stokes beam is made with the circular polarizer (CP) and a shear interferometer made of lens L2, the quartz plate (G) and $CCD_1$ camera. $CCD_2$ camera produces an image of the Stokes beam intensity.

The topological charge is separately analyzed for each circularly-polarized component of the CVB. The circular polarizer is made of a QWP with a fixed horizontal fast axis followed by a linear polarizer, whose transmission axis is set at 45º or -45º to transmit either right- or left-circular polarization states. A shearing interferometer made of a lens and a flat-parallel quartz plate (G) produces two copies of the beam laterally shifted by a small fraction of its diameter in order to make them mutually interfere. The topological charge of each circular component of the reflected beam can be identified from the fringe pattern recorded with a CCD camera. The presence of two dislocations in the fringe pattern, symmetrically placed on each side of the beam axis with the forks oriented in opposite directions is the signature of a topological charge, *m*, the value of which is determined by the number of branches in the fork minus one, and its sign by the orientation of the fork pattern [29].

Now, SBS is fundamentally a stochastic process triggered by random spontaneous Brillouin scattering events [4,5]. Here, for each prepared condition of the pump, Stokes CVBs with different topological charges are sometimes observed despite their lower gain. This is found to be particularly true when the gain of the competing modes is of similar magnitude, which happened near the poles of the HOPS, Cf. Table 1. Hence, for each condition, several snapshots, with sample size ranging from *N*=40 to *N*=60, are analyzed to reach statistically significant conclusions (Cf. ref. [18], section 2.2). However, for each condition, there is a dominant topological charge that occurs significantly more often.

Typical observations of the dominant topological charge are shown in Fig. 4. No conjugation of the OAM is found to take place at or near the equator of the HOPS. However, as the polar angle on the HOPS, $\theta$, decreases, the topological charge of each circular component shifts towards lower values. For instance, for the right-circularly polarized component, the dominant topological charge shifts from *m*=1 (no change) to *m*=0 and then to *m*=-1 (conjugation of OAM) as the $\theta$ value goes from 90º towards 0º, while the topological charge of the weaker left-circular polarization state shifts from *m*=-1 (no change) to *m*=-2 and then to *m*=-3.[2] These observations suggest that the CVB structure acquires a global phase factor exp(-2i$\varphi$) at small $\theta$ values such as:

$$\vec{E}_{s,-3,-1} \propto \exp(-2i\varphi)\left[\sin(\theta_s/2)\exp(-i\varphi) \quad \cos(\theta_s/2)\exp(i\varphi)\right]^T, \quad (21)$$

---

[2] This left-handed component however can only be measured down to about $\theta$=25º coordinate on the HOPS, below which it becomes too weak.

which reduces to a simple conjugation of the OAM when $\theta$ approaches zero, in agreement with our theoretical model (Cf. Table 2).

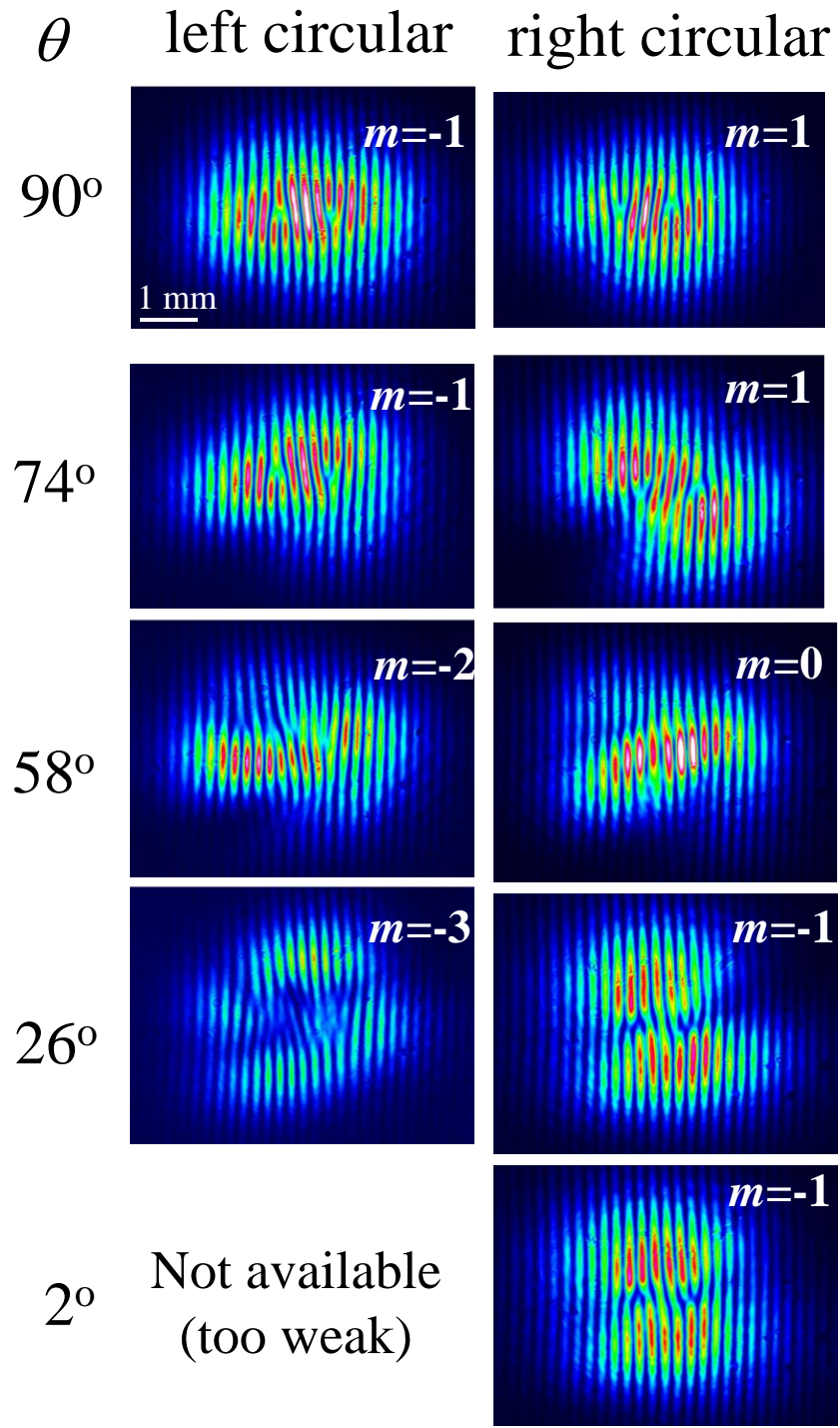

Fig. 4 Shearing interferometry snapshots taken with the CCD camera of the Stokes beam for incident pump CVBs of different $\theta$ values on the HOPS, analyzed into right- and left-circular components. The gradual shift of the OAM as $\theta$ approaches zero is clearly visible.

## IV. DISCUSSION AND CONCLUSION

In summary, using a theoretical model that considers both the inhomogeneous polarization distribution and the two OAM components of CVBs, we demonstrated that the optical conjugation of the topological charge of these OAM beams does not take place in SBS, except near the poles of the higher-order Poincaré sphere, where the beam approaches a scalar beam and phase conjugation of the OAM may take place. We traced this phenomenon to the fact that the Stokes beam has the tendency to acquire a polarization distribution similar to the pump beam. These results are summarized in Fig. 5. Near the equator, for oblong elliptical polarization, the Stokes vector mode with maximum gain is the same vector mode as the incident pump beam, e.g., Fig. 5a. As the beam becomes more scalar, e.g., Fig. 5b, the Stokes mode with maximum gain becomes that with topological charges $m=-1$ (i.e., the conjugate charge) for the dominant circularly polarized component and $m=-3$ for the weaker component. This corresponds to a precession of the phase with $\varphi$ in opposite directions for the pump and Stokes beams, without change in the polarization structure. Finally, for a pure scalar beam, the structure of the Stokes mode reduces to the conjugate topological charge, e.g., Fig. 5c. All these transformations gradually take place as $\theta$ changes.

We found that the eigenmodes of the Stokes beam were made of OAM modes of indices $n$-2 and $n$ with respectively left and right circular polarizations. This $\exp(i(n-2)\varphi)\hat{e}_L$ and $\exp(in\varphi)\hat{e}_R$ combination is to our knowledge the only one that maintains the axi-symmetrical structure of the beams, all other combination breaking this symmetry such as in the example shown in Fig. 2. These different modes merely differ by a helical phase factor that does not affect the polarization structure of the beam. Hence, all the Stokes vector eigenmodes are CVBs. Now, this does not mean that the polarization ellipse of the Stokes eigenmodes is identical to that of the pump. As a matter of fact, because the coefficients $\xi_{ij}$ in eq. (18) are not identical in general, the eigenvectors of the Stokes and pump beam, and hence the polarization states, are not identical in general. An example of the calculation of the eigenvectors of different modes is shown in ref. [18], section 1.4. It shows that the second component of the vector beam, which is the dominant component in the pump beam, becomes even more dominant in the Stokes wave and, hence, the Stokes beam tends to be more scalar than the pump beam.

Finally, it is important to point out some limitations in this model. First, it does not include the initiation of the SBS process by random spontaneous Brillouin scattering events. It also neglects the pump depletion and the potential interaction between Stokes modes [24]. It also ignores the radial structure of the Stokes modes, although our experiments did indicate that the wavefront curvature of the wavefront was indeed conjugated to the pump since the reflected Stokes beam retraced its path back with approximately the same divergence as the converging beam. Taking all these effects into account would require numerical simulations. Hence, this work should be viewed as a toy model that focuses on understanding why the SBS process does not conjugate the azimuthal phase structure of the CVB, except at the limit where such beam becomes a pure OAM mode where phase conjugation may take place.

Given the increasing significance of high-power CVBs for nonlinear optics [12-13], laser-matter interaction [14-15] and optical communications [30-31], the ability to correct optical aberrations incurred by the passage through an active or a turbulent medium is becoming increasingly important. SBS phase-conjugate mirrors are appealing to correct aberrations in a double-pass configuration. Understanding that the response of such mirror is affected by the entanglement between polarization and OAM in structured light is key to the successful use of SBS mirrors with these beams.

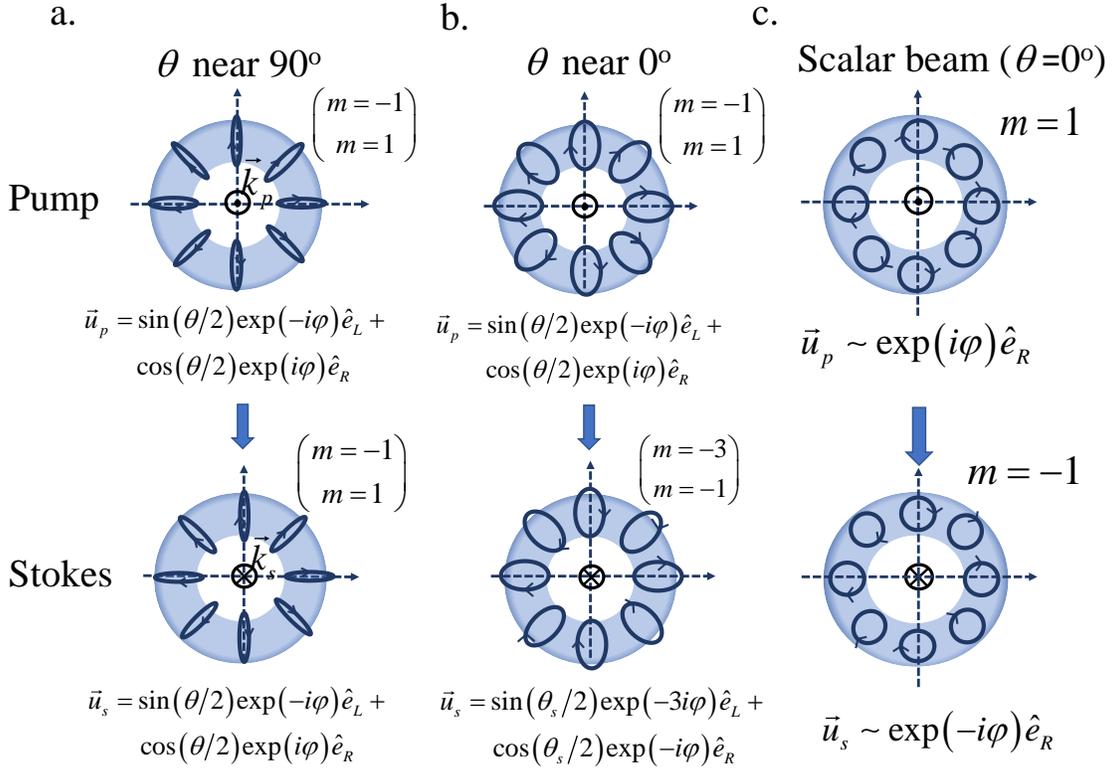

Fig. 5. Evolution of the state of the Stokes beam for CVBs of different polar angles $\theta$ on the HOPS.

a: Near the equator, the polarization distribution and the topological phase of the Stokes mode with highest gain remain unchanged with respect to the pump.

b: Towards the poles, the Stokes mode with highest gain acquires a global phase factor $\exp(-2i\varphi)$.

c: At the pole, the CVB reduces to a scalar beam and the situation observed in Fig. 5b reduces to the conjugation of the topological charge. The azimuthal phase distribution is indicated by the positioning of the arrow on each ellipse of polarization, which shows the helicity of the wavefront.

## ACKNOWLEDGMENT

The author acknowledges financial support from the Natural Sciences and Engineering Research Council of Canada Discovery grant program and from the Canada Foundation for Innovation.

End of the manuscript.